\documentclass[twocolumn,secnumarabic,floatfix,amssymb,amsmath,nofootinbib,tightenlines,showpacs,superscriptaddress,aps,prb]{revtex4}
\usepackage{amsfonts}
\usepackage{bm}%
\usepackage[colorlinks=true,linkcolor=blue,filecolor=blue,menucolor=yellow,urlcolor=blue,citecolor=blue,anchorcolor=blue]{hyperref}
\usepackage{graphicx}
\usepackage{dcolumn}
\usepackage{times}

\newcommand{\mun}{\widetilde{\mu}_N}
\newcommand{\sns}{S$\mid$N$\mid$S }
\newcommand{\sn}{S$\mid$N }

\newcommand{\etal}{\emph{et al.}}

\begin{document}

\title{Characteristic energies, transition temperatures, 
 and switching effects in clean 
\sns graphene 
nanostructures} 

\author{Klaus Halterman }
\email{klaus.halterman@navy.mil}
\affiliation{Michelson Lab, Physics
Division, Naval Air Warfare Center, China Lake, California 93555}
\author{Oriol T. Valls}
\email{otvalls@umn.edu}
\altaffiliation{Also at Minnesota Supercomputer Institute, University of Minnesota,
Minneapolis, Minnesota 55455}
\affiliation{School of Physics and Astronomy, University of Minnesota, 
Minneapolis, Minnesota 55455}
\author{Mohammad Alidoust}
\email{mohammad.alidoust@ntnu.no}
\affiliation{Department of Physics, Norwegian University of Science and
Technology, N-7491 Trondheim, Norway}
\altaffiliation{Also at Department of Physics, Faculty of Sciences, Unicersity
of Isfahan, Hezar Jerib Ave, Isfahan 81746-73441,Iran}

\date{\today}

\begin{abstract} 
We study proximity effects in clean nanoscale superconductor-normal metal-superconductor
(S$\mid$N$\mid$S) graphene
heterostructures using a self-consistent numerical
solution to the continuum Dirac Bogoliubov-de Gennes (DBdG) equations.
We obtain results for the pair  amplitude and the local density of states (DOS),
as a function of doping 
and of the geometrical parameters determining the width
of the structures.
The superconducting correlations are found to penetrate the 
normal graphene 
layers
even when there is extreme mismatch in the normal and
superconducting doping levels, 
where specular Andreev reflection dominates. 
The local DOS exhibits peculiar
features, which we discuss,
arising from the Dirac cone dispersion relation and
from the interplay between the superconducting and Thouless
energy scales. 
The corresponding 
characteristic energies emerge in
the form of resonant peaks in the local DOS, that
depend strongly on the doping level,
as does the energy gap, which
declines sharply as the relative difference in doping
between the S and N regions is reduced.
We also linearize the  DBdG  equations and develop
an essentially analytical method
that determines the critical temperature $T_c$ of an \sns 
nanostructure self-consistently.
We find that for S regions that occupy a fraction of the coherence length,
$T_c$ can undergo substantial variations as a function of the relative doping.
At finite temperatures and
by
manipulating the doping levels, the self consistent pair amplitudes reveal
dramatic transitions between a superconducting
and resistive normal
state of the structure.
Such behavior suggests the
possibility of using the proposed system
as a carbon-based superconducting switch,
turning superconductivity on or off
by tuning the relative doping levels.
\end{abstract}


\maketitle

\section{Introduction.}
The  successful development
of methods to create large samples of graphene,\cite{cite:novoselov1,cite:novoselov2} 
has been followed
by 
recent efforts  to exploit 
its high electron mobility\cite{cite:du0,cite:sarma} and 
the peculiar band structure\cite{cite:castroneto} 
associated with its two dimensionality. 
A number of graphene based devices have been subsequently proposed\cite{cite:book}, including
field effect transistors,
quantum information storage systems, optoelectronic devices, 
and nanoscale superconducting systems.
In particular, the observation of 
superconductivity in  graphene,\cite{silva,cite:heersche,cite:Du,cite:Jarillo,cite:shail} 
either through doping or by means of superconducting
contacts, 
has fueled research activity involving
proximity effects in
normal  (N) and  superconductor (S) graphene 
regions that are in close electrical 
contact.\cite{cite:beenakker2}
Indeed, the presence of superconducting correlations in graphene is 
remarkable considering that
undoped graphene in
isolation is inherently nonsuperconducting even at low temperatures.
Striking evidence of  the peculiarities of 
superconductivity in graphene 
was the observation of a Josephson supercurrent
induced by two superconducting electrodes in close contact 
with the graphene.\cite{cite:heersche} 
The tunneling conductance of a junction consisting 
of an insulating barrier between graphene and a superconductor  
should exhibit 
oscillations\cite{cite:bhat}
as a function of barrier strength, surprisingly peaking at finite values.
These unexpected effects
arise in large part from the hexagonal symmetry of graphene, which
generates
a relativistic-like  band structure\cite{cite:castroneto} near 
six points on the Fermi surface-the so called  Dirac points.
The low energy dispersion
near these points is  {\it linear},
and subsequently, the quasiparticles
are governed by 
a two dimensional massless Dirac-like equation.

Superconducting proximity effects in conventional heterostructures
consisting of a normal metal and superconductor have been
known for a very long time.\cite{cite:deush}
If the superconductor is coupled to
a graphene sheet,
where ``Dirac quasiparticles" are confined to a 2-D plane,
the leakage of superconductivity into graphene
should exhibit novel behavior.
Thus, studying superconducting proximity effects in graphene 
requires a careful and accurate determination of 
the pair correlations throughout the entire system. 
These are characterized by the pair potential $\Delta({\bm r})$, 
and the pair amplitude, $F({\bm r})$.
A proper delineation of the associated
proximity effects can only be achieved
through a self consistent calculation
of $\Delta({\bm r})$,
ensuring that the system's
lowest free energy state is found.
The resultant self consistent state
generally possesses nontrivial spatial inhomogeneity 
that can have important consequences for
quasiparticle bound states, interface bound states\cite{burset} 
and potential supercurrent flow. It is not surprising then that the 
frequently used step-function model for $\Delta({\bm r})$, while
satisfactory for length scales
much longer than the superconducting coherence length, $\xi_0$, can 
lead to erroneous results for small graphene structures where quantum scale oscillations
play a role.
For example, superconductor widths that are
on the same order as $\xi_0$ give rise to 
self consistent pair potentials that can vary over a significant
fraction of the total sample width. 
Moreover,  self consistency is crucial 
at finite temperatures, where
the superconducting correlations can have substantial decay near the interfaces.

The usual superconducting
proximity effect is governed by the mechanism
of Andreev reflection. This is the process where at the
interface, an electron with energy below
the superconducting energy gap is retro-reflected as a hole, transmitting a
Cooper pair into the superconductor. In graphene, the effectiveness of the
Andreev process depends in part on the relative doping in the S and N regions.
For electron doping
the Fermi level is shifted upwards, while for hole doping, it is shifted downwards 
relative to the Dirac point.
If the normal 
graphene layer is weakly doped, 
so that its Fermi level, $\mu_N$, is in absolute value
much smaller than that in the S region ($|\mu_N/\mu_S| \ll 1$),  
{\it specular} Andreev reflection becomes important. In this
process, the electron 
and hole belong to different bands.\cite{cite:beenakker1} 
Thus, despite large Fermi wavevector mismatch,
superconducting correlations can penetrate into the normal graphene region.
If on the other hand, $\mu_N/\mu_S \gg 1$, both the conventional and specular 
Andreev reflection processes
are suppressed and normal scattering drives the
quasiparticle trajectories.
The doping level clearly then has important 
consequences for any thermodynamic and
transport properties involving
superconducting graphene nanojunctions.

Besides doping effects, there are  geometrical issues to
contend with in finite S$\mid$N  nanojunctions: 
the electronic structure of
confined graphene can lead to a strong size dependence.\cite{cite:pono} 
Depending on the widths 
of the normal graphene and superconducting regions,  
there are various energy scales that 
can be difficult to disentangle.
If a \sns heterostructure has a thin middle
channel of width $d_N$ much smaller than $\xi_0$ (the
superconducting coherence length), 
the relevant low energy scale is the usual energy gap, $\Delta_0$.
This holds then for superconducting 
graphene\cite{cite:been2} which behaves, 
in this
case and in this respect, 
the same way as
conventional three-dimensional
materials.\cite{cite:heida,cite:nikolic} 
Short structures also result in critical currents that can
deviate from the simple harmonic form.\cite{cite:imry}
For wide 
middle layers, with $d_N \gg \xi_0$,
the Thouless energy, $E_T \equiv v_f/d_N$, ($v_f$ is the Fermi velocity) 
emerges as an
important energy scale.\cite{cite:been3}  
The Thouless energy in clean systems
gives rise to geometry dependent quantum phenomena 
that arise from 
the cumulative phase coherent effect
of
propagation and reflections from
the structure boundaries. This energy scale interacts then with
the $\Delta_0$ scale: 
for large  normal graphene widths, 
$E_T$ can be smaller than $\Delta_0$, and
the energy spectrum possess a Thouless gap for 
quasiparticles with energies less than 
a characteristic energy of order $E_T$. 
When $E_T$ is of the same order 
as $\Delta_0$, identifying the origin of
spectral anomalies can be difficult.
For long \sns heterostructures, 
the resultant  peaks in quasiparticle spectra
would be  (assuming a non self-consistent, step function 
form for $\Delta ({\bm r})$) 
located at energies proportional to
integer multiples of $E_T$.\cite{cite:been2} 
Self consistency can modify this result however as 
the pair potential deviates substantially from 
a simple step function model.
Moreover, when the doping amount changes,
this picture becomes complicated by changes
in quasiparticle bound states and density of states (DOS) 
due to the shifting of the
Fermi level. 

In this paper we use a fully self consistent framework to 
calculate the energy spectrum and pair amplitude in \sns
graphene heterostructures.
In addition to
the self consistent pair amplitude,
our accurate numerical 
diagonalization method allows us to  
investigate two important quantities that can be measured
experimentally and further clarify
proximity effects in graphene.
The first is the local density of states (DOS), which
can be measured directly with a scanning tunneling microscope.
The vanishing of the DOS at the Fermi energy in undoped graphene 
results in varying subgap bound states and minigaps\cite{burset2}
associated with the interplay between 
$E_T$ and $\Delta_0$.
To effectively characterize the local
electronic properties, we
examine the DOS,
in both the S and N regions.
The energy spectra reveal
conditions for fully gapped and 
gapless states in graphene \sns junctions.
For a given doping level,
the gap width and magnitude are shown to
diminish as $d_N$ increases. The
greatest variations are found to occur when $E_T < \Delta_0$.

The second experimentally observable quantity of interest 
is the critical temperature, $T_c$.
We study how $T_c$ varies as a function of doping levels and
of the geometrical parameters.
Our self consistent calculations find a nontrivial variation in
$T_c$ as a function of the relative doping levels, $\mu_N/\mu_S$.
The sensitivity of $T_c$ on the Fermi shifts depends strongly on the width of 
the outer S regions: 
very thin superconductors with $d_S < \xi_0$ reveal
the most drastic changes in $T_c$ for small increments in doping.
We show that, for particular ranges of $d_S$, $d_N$,
and temperature, 
a  \sns nanostructure can
act as a type of switch that
transitions between a
superconducting and resistive normal state
as the
ratio $\mu_N/\mu_S$ is varied, something that might be
done by using a modulated 
in-plane external electric field\cite{cite:novoselov1,ahn}.

Despite the importance of self-consistency, 
the only previous self-consistent works 
addressing proximity
effects in \sns  structures are Josephson
junction studies based on an extended Hubbard model 
with 
the superconductivity
arising either from doping\cite{cite:annica} or
from external contacts.\cite{cite:annica2,cite:annica3} 
It is the aim of this paper to present a method that
complements tight binding approaches
and provides a suitable
description for the critical temperature and applicable for 
a wide range of geometrical and coherence lengths. 
We achieve this goal by
numerically solving
the microscopic Dirac Bogoliubov-de Gennes (DBdG)
equations self consistently in the continuum regime. 
By retaining atomic length scales in the calculations,
we can accurately represent the important geometrical effects inherent to finite sized 
junctions.
The DBdG equations are ideal for inhomogeneous \sns heterostructures 
since they give directly the quasiparticle amplitudes and energies 
that characterize proximity effects. They are also appropriate for
clean systems such as graphene 
which has high electron mobility.
To investigate the potential 
of our \sns system as a superconducting graphene switch, 
we determine the critical temperature self consistently.
This is accomplished
by taking the full DBdG equations, 
and linearizing them via standard perturbative techniques.
We then arrive at an essentially analytical method that determines
the critical temperature as a function of geometrical parameters and doping levels.

\section{Method}
The geometry we study consists of a graphene sheet infinite in one
direction (that of the $y$ axis) and comprised of two doped  strips 
of superconducting material, each of width $d_S$, separated by a normal 
region of width $d_N$. 
We consider the pairing in the S regions
to be conventional $s$-wave.
The methods we use to self-consistently
diagonalize the mean field single-band Hamiltonian are extensions of those 
previously employed\cite{cite:hv02a,cite:hv04a,cite:hv05,cite:bhv07} to study proximity
effects in ordinary three dimensional materials, but important changes
have to be made to take into account the reduced dimensionality and the
peculiar band structure of graphene. These changes are the focus of
the discussion below.

Our starting point in this case is the 
Dirac Bogoliubov-de Gennes (DBdG) 
equations which govern 
the quasiparticle spectrum of graphene.
In the absence of 
magnetic effects, 
the DBdG equations for the two valleys $K (+)$ and $K' (-)$ are \cite{cite:beenakker1}:
\begin{align}
&\begin{pmatrix}
{\cal H}^\pm-\mu \hat{I}&\Delta \hat{I} \\
\Delta^*\hat{I}&-({\cal H}^\pm-\mu \hat{I})
\end{pmatrix}
\begin{pmatrix}
\Psi^\pm_{u,n}\\ \Psi^\mp_{v,n}
\end{pmatrix}
=\epsilon_n
\begin{pmatrix}
\Psi^\pm_{u,n}\\ \Psi^\mp_{v,n}
\end{pmatrix}\label{eq:hamiltonian1},
\end{align}
The Dirac Hamiltonian, ${\cal H}^\pm$, is given compactly by 
$ {\cal H}^\pm=v_f (\sigma_x p_x \pm \sigma_y p_y)$,
in which $\sigma_i$ are the $2\times2$ Pauli matrices acting in sublattice space,
$\hat{I}$ is the identity matrix, $v_f$ is the (energy independent) Fermi velocity 
in graphene,
and $\mu$ is the chemical potential.
This quantity vanishes
in the undoped case but not in the presence of doping. Here
we will take $\mu(x)$ to be a piecewise
constant: a fixed positive number $\mu_S$ in the $S$ region
and a variable value 
$\mu_N$ in the $N$ regions.
We consider the case of relatively large doping in the $S$ regions,
so that we can assume smooth interfaces.\cite{cite:beenakker1,cite:annica2}
We will take these interfaces in the direction
of constant $x$. 
We have defined $\Psi^+_{u,n} \equiv (u^n_{A,K},u^n_{B,K})^T$,
$\Psi^-_{u,n} \equiv (u^n_{A,K'},u^n_{B,K'})^T$, $\Psi^+_{v,n} \equiv (v^n_{A,K},v^n_{B,K})^T$,
and $\Psi^-_{v,n} \equiv (v^n_{A,K'},v^n_{B,K'})^T$. The $A,B$ labels
denote the 
two sublattices that arise from the honeycomb lattice structure.

The notation in the Hamiltonian implies that the 
Pauli matrices act on the pseudospin of the quasiparticles,
mapping the usual spin into
the projection of the wavefunction onto sublattice A or B.
Since the valleys are degenerate ($K$ and $-K'$
are equivalent), we need only solve for either $\cal{H}^{+}$
or $\cal{H}^{-}$. 
Assuming the first choice, we define the four component
vector $\Psi_n \equiv (\Psi^+_{u,n},\Psi^-_{v,n})$. 

The pair potential $\Delta$ couples electrons in a given valley with
the  hole excitations in the other valley. 
In terms of the wavefunctions and energies obtained from Eq.~(\ref{eq:hamiltonian1}),
this coupling leads to the self-consistency condition,
\begin{align}
\label{eq:del}
\Delta(x) =  \frac{g}{2}{\sum_n} 
\bigl[u^n_{A,K} v^{n*}_{A,K'}
 + u^n_{B,K}v^{n*}_{B,K'}
\bigr]
\tanh\Bigl(\frac{\epsilon_n}{2T}\Bigr), \,
\end{align}
where the superconducting coupling parameter, $g$, is a positive constant 
in the intrinsically superconducting
regions and zero elsewhere. 
The sum is over all  energy eigenstates
within the Brillouin zone whose energy, referred to $\mu_S$, 
is smaller than or equal to a
characteristic energy cutoff, $\omega_c$.
It is to be interpreted as
$\sum_n \rightarrow 1/(2\pi) \int dk_y \sum_q$,
where $k_y$ is the transverse momentum, and $q$ a  longitudinal index.
The singlet pairing only occurs from opposite valleys,
to maintain time-reversal symmetry \cite{cite:beenakker1,cite:beenakker2}.

The experimentally important local DOS, $N(x,\epsilon)$,  is given by
\begin{align}
\label{dos}
N(x,\epsilon)=-
{\sum_{n,\alpha,\beta}}    
\Bigl[|u^n_{\alpha,\beta}|^2 f^\prime(\epsilon-\epsilon_n)
+|v^n_{\alpha,\beta}|^2f^\prime(\epsilon+\epsilon_n)\Bigr],
\end{align}
where $\alpha$ equals $A$ or $B$,  $\beta$ can be either $K$ or $K'$,
and $f^\prime$ is the derivative of the Fermi function.
One can integrate $N(x,\epsilon)$ over any suitable range
of $x$ to obtain the average DOS in a certain region. 

We now take advantage 
of the translational invariance along $y$ 
by writing $\Psi_n(x,y) \equiv e^{i k_y y} \Phi_n (x)$.
Introducing the notation $\Phi^T_n(x) \equiv ( s_n(x),t_n(x),w_n(x),z_n(x) )^T$
where the functions in the parenthesis correspond to valley and sublattice
indices in the same order as for the previously defined 
$\Psi_n$, we can rewrite  
the  DBdG equation Eq.~(\ref{eq:hamiltonian1}) as,
\begin{align}
\label{new}
&\begin{pmatrix}
-\mu&\pi^{\dag}_{+}&\Delta&0 \\
\pi^{\dag}_{-}&-\mu&0&\Delta \\
\Delta^*&0&\mu&\pi_{+} \\
0&\Delta^*&\pi_{-}&\mu
\end{pmatrix}
\begin{pmatrix} 
s_n\\t_n\\w_n\\z_n
\end{pmatrix}
=\epsilon_n
\begin{pmatrix}
s_n\\t_n\\w_n\\z_n
\end{pmatrix},
\end{align}
where we define $\pi_{\pm}\equiv i  v_f (\partial_x\pm k_y)$,
(we use $\hbar=1$ and $k_B = 1$ throughout this paper).  
%

Next we expand the quasiparticle wavefunctions via,
\begin{align}
\label{expand}
\Phi_n (x) = \sum_{q=1}^N
{\bm c}_{n,q}
\phi_q (x),
\end{align}
where the ${\bm c}_{n,q} \equiv
(s_{n,q},t_{n,q},w_{n,q},z_{n,q})^T$,
are the
expansion coefficients, and the $\phi_q (x)$ is a set of $N$ basis
functions, where $N$ must be sufficiently large.\cite{cite:hv02a,cite:hv04a} 
We take, consistent with the boundary conditions, 
$\phi_q (x) \equiv \sqrt{2/d} \sin(k_q x)$ in which $d=2d_S+d_N$,
and $k_q = q \pi/d$ is the quantized wavenumber. These basis functions
are not eigenstates of the 
normal Hamiltonian. Therefore
there are $\pi^{\pm}$ off diagonal terms. 
These introduce some computational challenges that
result in a larger value of  $N$  being  
required than in the three dimensional cases. 
Considering  then  each row of Eq.~(\ref{new}), we perform the
following somewhat lengthy but elementary steps. First,
we insert the expansion Eq.~(\ref{expand}) into Eq.~(\ref{new}).
Next, we
multiply each term by $\phi_{q'}$ and 
integrate the variable $x$ over the range $0\le x \le d$,
taking into account properly the stepwise $x$ dependence of $\mu$. 
Finally, we choose $\mu_S$ as our unit of energy (recall
that we are dealing with a strongly doped system so this
quantity does not vanish) and divide
through by $\mu_S$.

Taking the same
steps for the rest of the matrix,
we end up with the following $4N\times4N$
matrix equation:
\begin{align}
\label{eq:mat}
{\bf M} {\bm a}_n = \widetilde{\epsilon}_n {\bm a}_n,
\end{align}
where
\begin{align}
{\bf M} =
&\begin{pmatrix}
{\cal A}&{\cal B}-i \widetilde{k}_y {\mathbb{I}}&{\cal D}&{\mathbb O} \\
{\cal{B}} +i \widetilde{k}_y {\mathbb I}&{\cal A}&{\mathbb O}&{\cal D} \\
{\cal D}^*&{\mathbb O}&-{\cal A}&-({\cal B}-i \widetilde{k}_y {\mathbb I})\\
{\mathbb O}&{\cal D}^*&-({\cal B}+i \widetilde{k}_y {\mathbb I})&-{\cal A}
\end{pmatrix},
\label{eq:bogo4}
\end{align}
and the vector, ${\bm a}_n$, contains the expansion coefficients,
\begin{align}
{\bm a}_n \equiv (
&s_{n,1},\cdots,
s_{n,N},t_{n,1},
\cdots,t_{n,N}, \nonumber \\
&w_{n,1},
\cdots,w_{n,N},
z_{n,1},\cdots,z_{n,N} )^T.
\end{align}
Here, ${\mathbb I}$ and ${\mathbb O}$ are
unit and zero  matrices of rank $N$ respectively. Consistent with
our choice of energy units, we now define tilded  dimensionless
energies $\widetilde{\mu}_N \equiv \mu_N/\mu_S$ and
$\widetilde{\epsilon}_n \equiv \epsilon_n/\mu_S$.
We choose also to measure our wavectors in units of $k_{FS}$
defined by the relation $k_{FS} \equiv \mu_S/v_f$, 
thus e.g.
$\widetilde{k}_y\equiv k_y/k_{FS}$ and then have for the
remaining elements in Eq.~(\ref{eq:bogo4}):
\begin{align}
&\nonumber{\cal A}_{q,q'} = {\cal K}_{q+q'}(d_S)-{\cal K}_{q-q'}(d_S)
-\widetilde{\mu}_N[{\cal K}_{q-q'}(d_S+d_N)
\\&\nonumber-{\cal K}_{q+q'}(d_S+d_N)
-{\cal K}_{q-q'}(d_S)  +{\cal K}_{q+q'}(d_S)]\\&+
{\cal K}_{q-q'}(d_S+d_N)-{\cal K}_{q+q'}(d_S+d_N), \quad q\neq q',
\end{align}
\begin{align}
{\cal A}_{q,q} = {\cal K}_{2 q}(d_S)-\dfrac{d_S}{d}
-\widetilde{\mu}_N[\dfrac{d_N}{d}+{\cal K}_{2 q}(d_S)
\nonumber \\-{\cal K}_{2 q}(d_S+d_N)] -
\dfrac{d_S}{d}-{\cal K}_{2 q}(d_S+d_N),
\end{align}
\begin{align}
{\cal B}_{q,q'} =  \dfrac{ 2 i  q' q}{k_{FS} d} \Biggl (\dfrac{-1 + (-1)^{q+q'}}{q^2-q'^2} \Biggr), \quad q\neq q',
\end{align}
and ${\cal B}_{q,q} = 0$.
Here, 
${\cal K}_n(x) \equiv \sin(n \pi x/d)/(n\pi)$.
Finally:
\begin{align}
{\cal D}_{q,q'} = \dfrac{2}{d}\dfrac{\Delta_0}{\mu_S}\int_0^d dx \sin(k_q x) (\Delta(x)/\Delta_0) \sin(k_q' x),
\end{align}
where $\Delta_0$ is the order parameter in bulk $S$ material at $T=0$.
The self-consistency relation Eq.~(\ref{eq:del}) can now be written as, 
\begin{align}  
\label{del2} 
\Delta(x)/\Delta_0 &=  4\lambda\Bigl(\dfrac{\xi_0}{d}\Bigr){\int_0^{k_c} 
d\widetilde{k}_y\sum_n \sum_{q,q'}  }  
\bigl(s_{n,q} w_{n,q'}^* \\ \nonumber
&+t_{n,q} z_{n,q'}^*\bigr)\sin(k_q x)\sin(k_q' x) 
\tanh\Bigl(\frac{\epsilon_n}{2T}\Bigr),
\end{align} 
where\cite{cite:sonin1} $\xi_0=v_f/\Delta_0$, $k_c$ is the $\widetilde{k}_y$ 
cutoff corresponding to those states specified below Eq.~(\ref{eq:del}),
and $\lambda$ is the {\it dimensionless}
coupling constant which we define as 
$\lambda \equiv g \mu_S /2\pi v_f^2$. Since the DOS
for bulk $S$ material in its normal state 
is $N_0(\epsilon) = 2 \epsilon/(\pi v_f^2)$, we have 
$\lambda=g N_0(\mu_s)/4\pi$. 

To perform our calculations, we must solve Eq.~(\ref{eq:mat}) together with the
self consistency condition Eq.~(\ref{del2}). 
When $\widetilde{\omega}_c \equiv  \omega_c/\mu_S$ satisfies 
$ \widetilde{\omega_c}\le 1$, 
we find for the bulk case,
\begin{align}
\label{lam}
\lambda^{-1}= {\rm arcsinh}(\omega_c/\Delta_0), 
\end{align}
while if $\widetilde{\omega}_c \ge 1$, then \cite{cite:sonin1} 
\begin{align}
\label{lam1}
\lambda^{-1}=
\sqrt{\widetilde{\Delta}_0 ^2+\widetilde{\omega}_c^2}- \sqrt{\widetilde{\Delta}_0^2+1}+{\rm arcsinh}(1/\widetilde{\Delta}_0 ).
\end{align}
where  
$\widetilde{\Delta}_0 \equiv \Delta_0/\mu_S$. 
To achieve self consistence, we start with an initial guess 
for $\Delta(x)$ and once all of the eigenfunctions
and eigenenergies have been determined, we
calculate a new $\Delta(x)$ from Eq.(\ref{del2}) and
iterate this process until the relative difference 
between successive $\Delta(x)$ 
is less than $10^{-4}$. 

We 
also determine 
the critical temperature, $T_c$,
semi-analytically
as a function of the geometrical and doping parameters.
To find $T_c$,
the self-consistency equation can be linearized\cite{cite:bhv07}  near the
transition, leading
to the form 
\begin{align}
\label{tc}
\Delta_i=\sum_q J_{iq}\Delta_q, 
\end{align}
where the $\Delta_i$ are expansion coefficients 
of the  pair potential (Eqs.~(\ref{eq:del}) and (\ref{del2})) in 
our basis and the  $J_{iq}$ are the appropriate matrix elements with respect to the same basis, 
as obtained from the
linearization procedure. These  matrix elements can be written 
as $J_{iq} \equiv (J_{iq}^u+J_{iq}^v)/2$, where,
\begin{align}
J_{i q}^u&=
\gamma
\int d \widetilde{k}_y
\sum_n^{N_D} \Biggl[ \tanh\biggl(\frac{ \widetilde{\epsilon}_n^{u,0}}{2{T}}\biggr)
 \sum_m^N \frac{{\cal F}_{qnm} {\cal F}_{inm}}
 {\widetilde{\epsilon}_n^{u,0}-\widetilde{\epsilon}_m^{v,0}} \Biggr ], \\
J_{i q}^v&=
\gamma \int d  \widetilde{k}_y
\sum_n^{N_D} \Biggl [ \tanh\biggl
( \frac{\widetilde{\epsilon}_n^{v,0}}{2{T}}\biggr)
 \sum_m^N \frac{{\cal F}_{qmn} {\cal F}_{imn}}
 {\widetilde{\epsilon}_n^{v,0}-\widetilde{\epsilon}_m^{u,0}} \Biggr ].
\label{matfinal2}
\end{align}
Here $\gamma =\lambda/(2\pi^2 k_{FS} d)$, with $\lambda$ the {\it dimensionless} 
superconducting 
coupling constant introduced above. 
The eigenenergies, $\epsilon_n^{u(v),0}$, are the unperturbed particle (hole)
energies (found by setting ${\Delta} = 0$ in Eq.~(\ref{new})), 
and $N_D$ denotes that the sum is
cut off at energies beyond
the $\omega_c$ frequency. 
We also have,
\begin{align}
{\cal F}_{qnm}\equiv \pi \sqrt{2 d} \sum^N_{p,r} {\cal K}_{q p r}
(s_{n r} w^*_{m p} +t_{n r} z_{m p}^* ),
\end{align}
where the correlation factor, ${\cal K}_{q p r}$, is written,
\begin{align}
{\cal K}_{q p r} \equiv (2/d)^{3/2}  \int_0^d dz \Theta(z)\sin(k_q z) \sin (k_p z) \sin (k_r z). 
\end{align}
Here 
we define
$\Theta(z)$ to be unity in the superconducting regions and vanish in the normal
ones.  
The determination of $T_c$  involves
calculating the eigenvalues of matrix $J_{iq}$ for each 
temperature value in the range of interest, 
and the highest temperature for which the largest eigenvalue 
of the matrix $J_{iq}$ is unity\cite{ad,lv}
corresponds to $T_c$. 
This linearization route is much more efficient at calculating $T_c$  
than solving the full DBdG equations near $T_c$, which can
often be very difficult due to the large number of iterations involved 
in the self consistency process.

\begin{figure*}
\includegraphics[width=7 in]{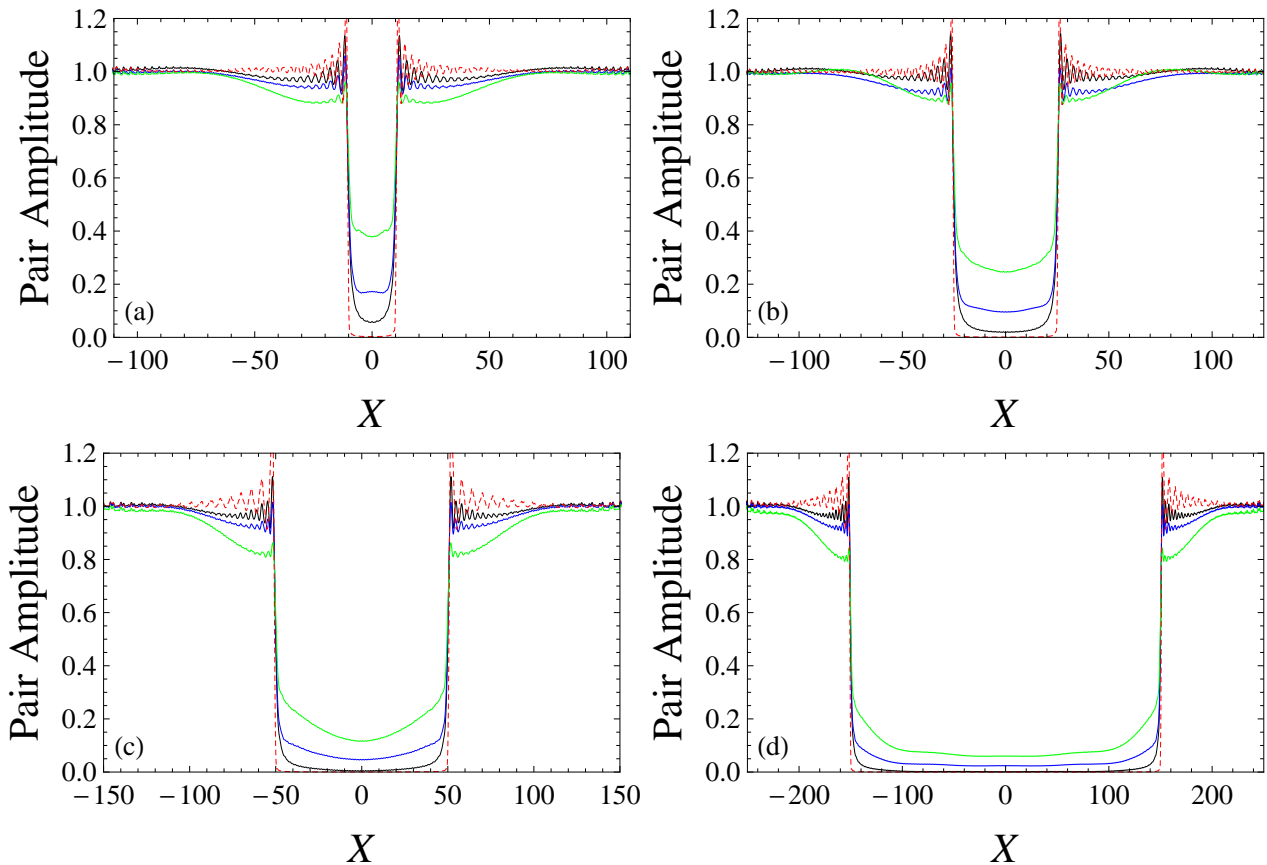}
\caption{(Color online). Normalized pair amplitude vs $X$ (see text)
at $T=0$ for an \sns heterostructure.
The S portions have a dimensionless width (see text) of $D_S = 150$ each. 
For clarity, the outermost parts of the sample are not shown.  
Each panel corresponds to
a different normalized normal 
graphene width of (a) $D_N = 20$, (b) $D_N = 50$, (c) $D_N = 100$, and (d) $D_N = 300$.
For each case, four curves representing different doping levels
are shown.
From top to bottom in the central N region (green, blue,
black, red), these curves correspond to 
$\widetilde{\mu}_N = 0.5, 0.2, 0$, and $10$.
}
\label{f1} 
\end{figure*}

\begin{figure*}
\includegraphics[width=7in]{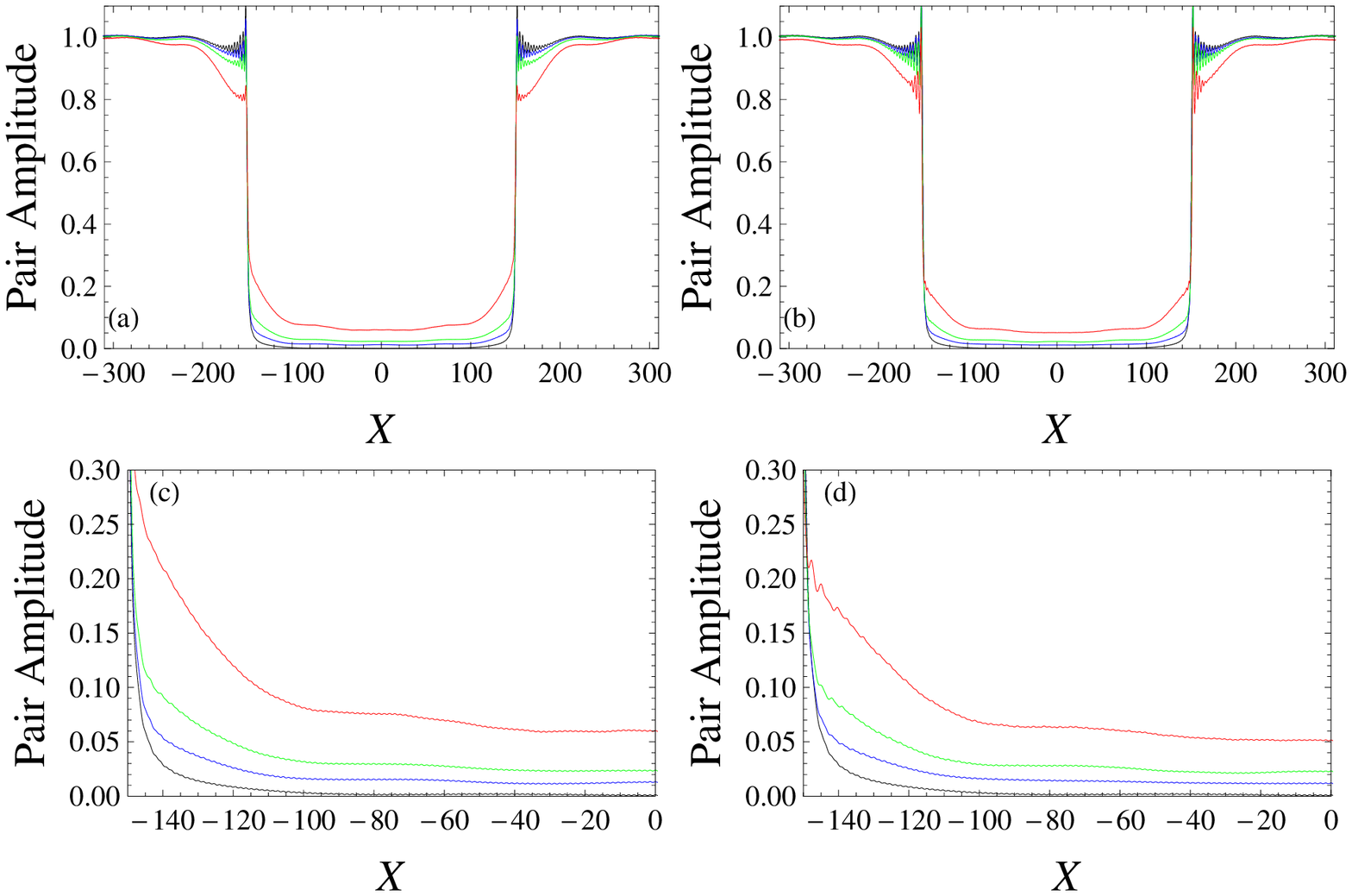} 
\caption{(Color online). Normalized pair amplitude 
versus $X$, as in Fig.~\ref{f1}. 
The dimensionless geometrical quantities $D_N$ and $D_S$ are now $D_N = 300$ and $D_S=300$,
so that the \sn boundaries are at $|X| = 150$.
The top
set of panels depicts most of the \sns structure 
and the bottom set of panels are magnifications
of part of the N region near the interface for the panel above.
Panels (a) and (c)
are  for electron doped N  with curves shown, from top
to bottom at $X=0$, for $\mun = 0.5,0.2$, and $0.1$ (red, green
and blue), while panels (b) and (d)
are for hole doped N with $\mun = -0.1,-0.2$, and $-0.5$
with a similar scheme. In both cases results for $\mun =0$ (black) lowest
curve at $X=0$, are shown 
for reference.
}
\label{f2} 
\end{figure*}
\section{Results and Discussion}
In this section,
we present our self consistent
results for the pair amplitude, local DOS, and critical temperature,
for a broad range of widths and relative doping levels.
We consider the S regions to be electron-doped, corresponding to $\mu_S > 0$, 
while the normal graphene can be either electron-doped ($\mu_N>0$) 
or hole-doped ($\mu_N<0$).
All lengths are scaled in units
of the Fermi wave vector $k_{FS}$, 
and we define the relative dimensionless
coordinate $X \equiv k_{FS}(x-d/2)$,
so that $X=0$ is
at the center of the structure. 
When considering 
thermal effects, all 
quantities involving the
temperature, $T$, are 
scaled by  $T_0$,
the transition temperature for the bulk superconducting material. 
Our input parameters are,
besides the geometrical lengths, which are given 
in dimensionless form as $D_S \equiv k_{FS}d_S$ 
and $D_N \equiv k_{FS}d_N$, the value of $\widetilde{\mu}_N$, and that of the 
dimensionless coherence length $\Xi_0 \equiv k_{FS} \xi_0 = 100$. 
From the
latter, and using a fixed value of $\widetilde{\omega}_c=0.04$ 
(for $\omega_c < 1$, 
results are only weakly sensitive 
to $\omega_c$) we obtain $\lambda$ via 
Eq.~(\ref{lam}). 

We will first consider our results for the
self-consistent  
normalized Cooper {\it pair amplitude} 
$F(x) \equiv (1/\lambda) \Delta(x)/\Delta_0$,
which reveals the superconducting
correlations throughout the entire \sns system. Some of our results
for $F(X)$ as a function of dimensionless distance $X$  are shown in 
Fig.~\ref{f1} and Fig.~\ref{f2}.
In Fig.~\ref{f1},
$F(X)$ is shown,
in each panel, for a different value of $D_N$,
and  all at the same $D_S=150$. 
Several (all positive) values of the relative doping parameter $\mun$,  
corresponding to electron
doping, are shown in this figure:
the undoped N case is also shown 
for comparison. 
We see that the proximity effect depends strongly on the 
relative doping $\mun$ via the mismatch it
reflects (when this
quantity is unity, there is no mismatch).
Moderate doping in the N region
allows for very appreciable pairing correlations in the normal graphene, even
when $D_N > \Xi_0$ (right bottom panel), as one can see e.g. in the
$\mun=0.5$ ((green) highest curves at $X=0$) 
results shown, while depletion of $F(X)$ in the S regions
extends in this case to distances longer than the
correlation length. On the other hand, when the 
mismatch in the Fermi shifts is extreme (as in the $\mun = 0$ ((black)
solid curve, third from top at $X=0$) and 
$\mun = 10$ ((red) dashed curves) cases
shown, we see that the proximity effect is much weaker. The 
(blue) curves, second highest at $X=0$ corresponding
to $\mun=0.2$, show intermediate behavior. 
Further
examination of the results in this figure for large mismatch ($\mun=0$) 
reveal that specular Andreev reflection  allows 
somewhat more readily for the penetration of
correlations in the normal graphene than large
mismatch in the opposite direction: comparing the $\mun = 0$  to the
$\mun = 10$ results, 
the self consistent state in the later case 
shows less superconducting correlations in the normal graphene
than in the former case, and a correspondingly smaller 
depletion in the S layers. 
The sequence of panels, moreover, illustrates that increasing the N graphene widths
always results (at the same value of $\mun$) in greater 
superconductivity depletion in the S regions near the interfaces
due to leakage into the N layer. The smaller the Fermi level mismatch,
the greater this effect.
For more confining N regions (smaller $D_N$),
the pair correlations decay in in the normal layer over a smaller width
and thus
the two superconductor portions of the sample 
are more strongly coupled.

\begin{figure*} 
\includegraphics[width=7in]{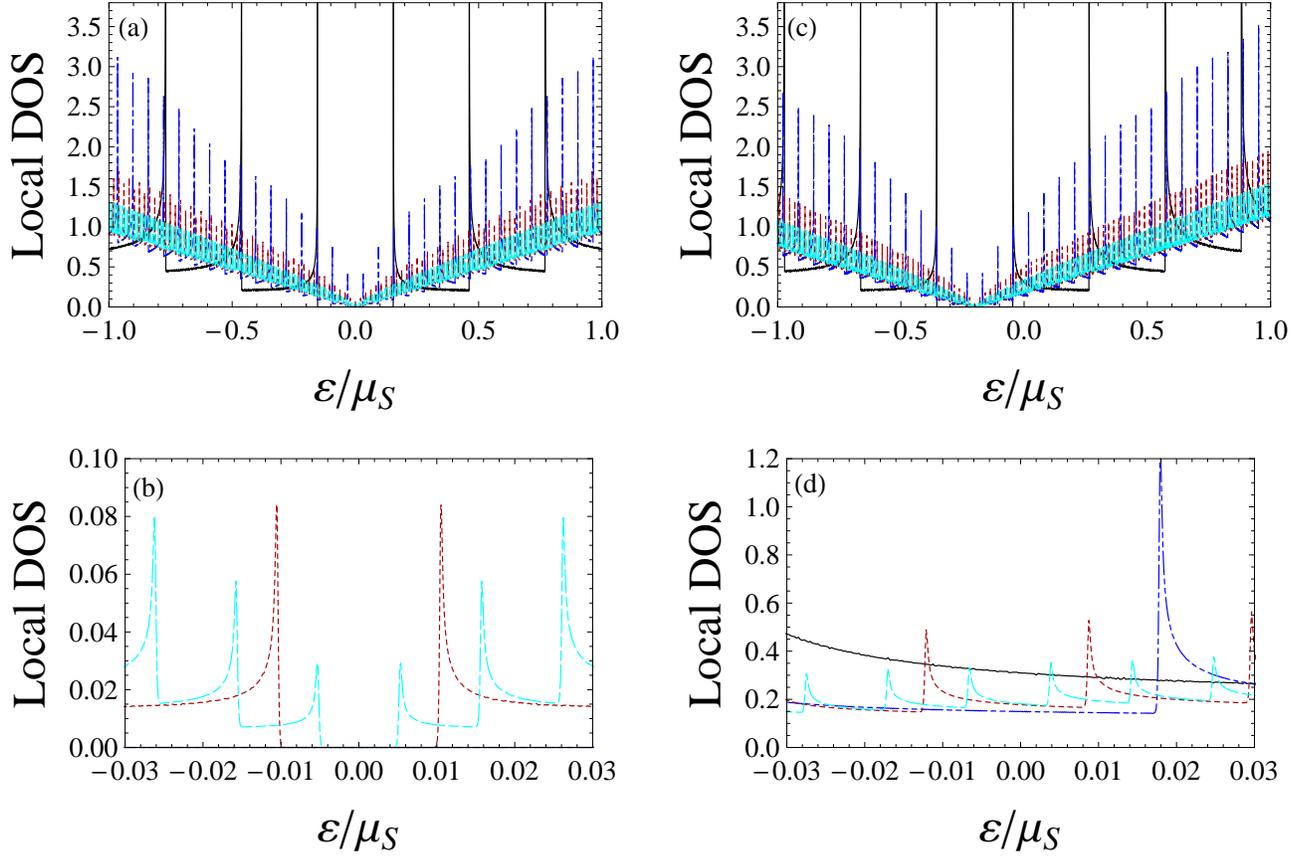}
\caption{(Color online). Local density of states (DOS)
for a normal, finite width, graphene 
layer that has (a) zero doping ($\mun = 0$), 
and (c) moderate doping ($\mun = 0.2$). The normalizations of DOS and 
energies are chosen
(see text) so that for an infinite width layer the plots would be
straight lines of slope $\pm 1$. Four different widths are illustrated, 
in order of decreasing height of the 
main peaks:
$D_N = 20$ (black), $D_N = 100$ (blue), $D_N = 300$ (red), and $D_N = 600$ (cyan).
Each bottom panel is a magnification of the results above it, 
over a narrower energy range. The main peaks are related to the
Thouless scale, see Eq.~(\ref{thou}) and discussion below it.  
}
\label{f30} 
\end{figure*}

\begin{figure*}
\includegraphics[width=7in]{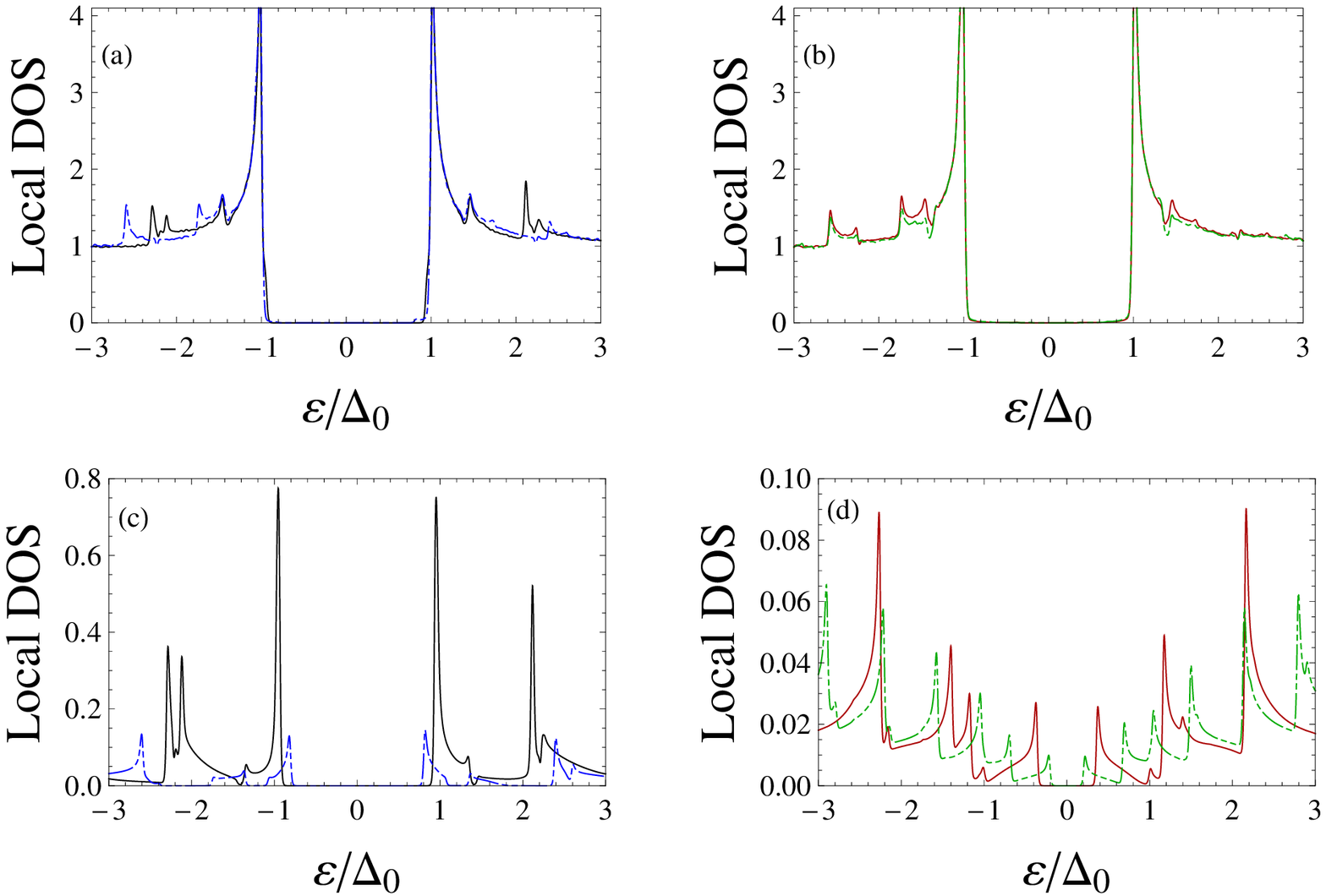}
\caption{(Color online). Local density of states (DOS)
(normalized as explained in the text)
for an \sns 
system with
highly doped S layers ($\mun = 0$).
Results are shown for the S
region (panels (a) and (b)) and
in the N region ((c) and (d)).
In (a) and (c), we have $D_N = \Xi_0/5$ (solid  (black) curves)
and $D_N = \Xi_0$ ((blue) dashed curves), while in (b) and (d)
the N layers are larger:
$D_N = 3 \Xi_0$ (solid (red) curves) and $D_N = 6 \Xi_0$ (dashed (green)
curves).
}
\label{f3} 
\end{figure*}

To better illustrate the slow 
decay of of the amplitude $F(X)$ in the normal graphene region
we display  in Fig.~\ref{f2}, results  for this
normalized quantity obtained  for a 
much larger system with $D_N,D_S > \xi_0$. We show there also 
results for  a broader range of doping levels in N.
The top set of panels shows a 
global view of the correlations
in the \sns structure, similar to that shown in Fig.~\ref{f1}, 
while in the bottom set of panels
are closeups of the normal graphene region near the interface.
The electron doped (left panels) cases
are nearly identical to the hole doped (right panels)
ones, except at smaller mismatch. The bottom panels
allow for a more detailed examination of the behavior
near the interface. We note that, if the mismatch is not
large, penetration of the Cooper pairs over a distance clearly much
larger than the correlation length occurs, and that depletion
in the S regions occurs also in the same scale. On the
other hand, one can see in both this and the previous figure that
the transition 
between the depleted S region and the weakly
proximity-influenced N region is very abrupt: there
are in effect two length scales, one related to the
depletion and penetration, which can be rather longer than $\xi_0$, and
another, very short scale, over which the small values
of $F(X)$ in N transition sharply to the depleted, but much larger,
values in S. 
This is  in contrast to what occurs in the standard proximity
effect in ordinary bilayer materials, which is characterized by a single length scale. 

\begin{figure*}
\centering
\includegraphics[width=7in]{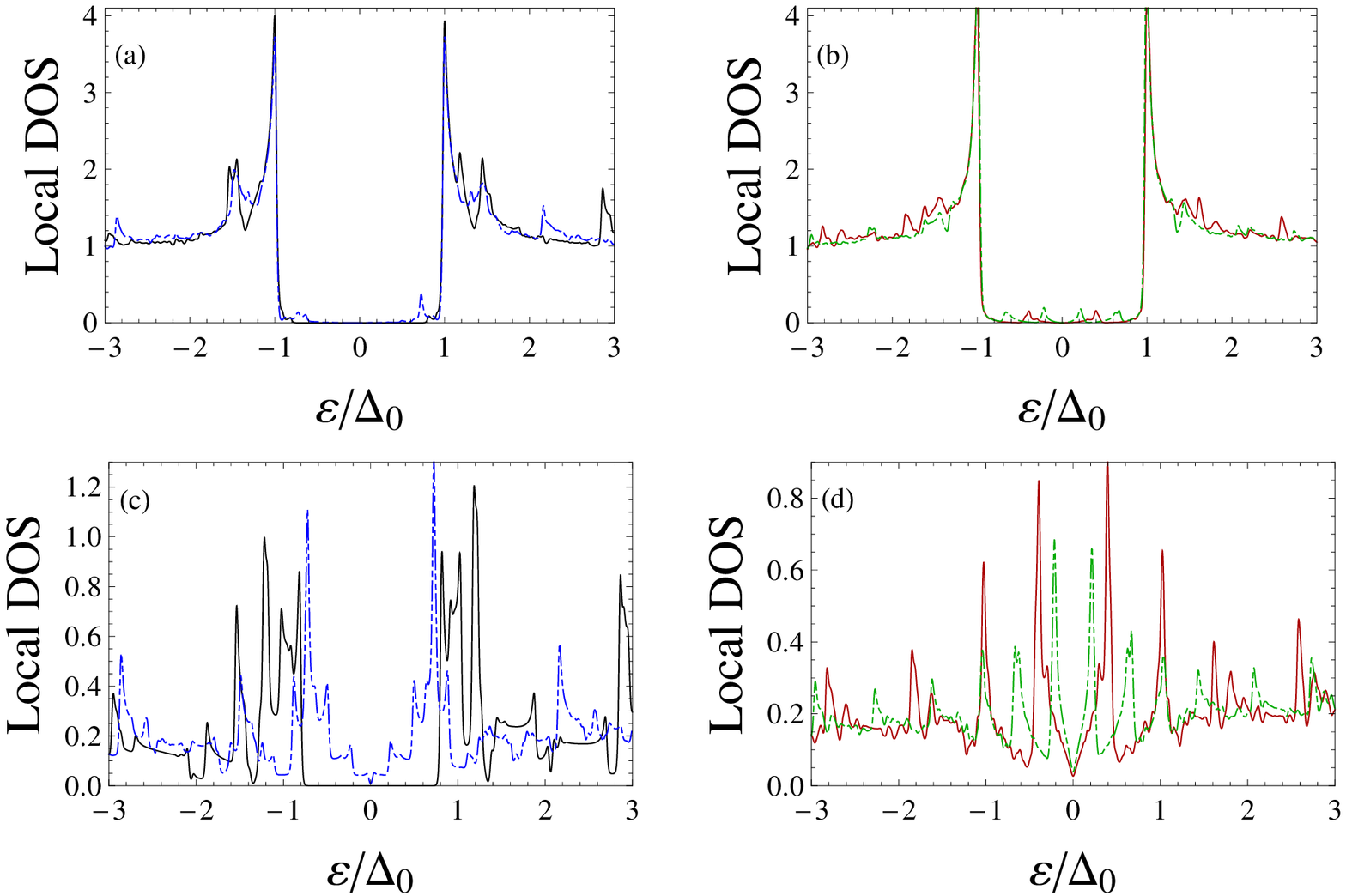}
\caption{(Color online). Local density of states 
(plotted and normalized as in Fig.~\ref{f3})
for an \sns 
system with
moderately doped S layers ($\mun = 0.2$).
The panel arrangement, curve
(color and) structure, and all other parameters
are the same as that in Fig.~\ref{f3}. See text
for discussion and comparison
}
\label{f4} 
\end{figure*}

Before discussing the local density of states
for \sns structures,
it is illuminating to first investigate the DOS and characteristic energies
for  
pure graphene nanolayers, which we do by setting $D_s=0$. 
In the absence of other materials, and hence also of proximity effects,
the DOS in this case is essentially independent of position, and
thus
it is appropriate to spatially
average Eq.~(\ref{dos}) over the entire sample, which (after using
the normalization condition for the
quasiparticle amplitudes) yields
a simplified normalized DOS, 
\begin{align}
\label{dos3}
\dfrac{N(\epsilon)}{N_0(\mu_S)}  
&=
\frac{v_f}{2 T d }\sum_n \int_0^{k_c} d\widetilde{k}_y  
\Bigl[ \cosh^{-2}
\Bigl(\dfrac{\epsilon-\epsilon_n}{2T}\Bigr) \nonumber \\
&+ \cosh^{-2}
\Bigl(\dfrac{\epsilon+\epsilon_n}{2T}\Bigr)
\Bigr],
\end{align}
where $T$ is the temperature,
and $N_0(\epsilon)$ is introduced 
below Eq.~(\ref{del2}). In this case, with only 
normal material present, it must be understood
that  $N_0(\mu_s)$ is just an arbitrary, but convenient, 
normalization. 
Note that after setting $D_S=0$ and hence $\Delta \equiv 0$, 
no iteration for self consistency is needed and  
only the eigenvalue
spectra needs to be determined when performing the 
diagonalization of the matrix in Eq.~(\ref{new}).
To achieve the required energy resolution
for the results to follow,
integrals over $k_y$ are numerically evaluated by transforming them into a sum 
over 5000 transverse modes.
Also, in order to better discern the relevant DOS features,
we 
consider 
the  low temperature limit (see below). 
For finite width graphene sheets, 
the coherent superposition of standing waves 
determines the  Thouless energy scale,\cite{cite:been3} 
\begin{align} 
\label{thou}
E_T= \dfrac{v_f }{d_N}. 
\end{align}
The Thouless energy scale reveals itself in the form of
peaks in the quasiparticle spectra that, in this simple geometry 
repeat at odd integer multiples of $E_c = \pi E_T$. 
We have then $E_c/\mu_S= \pi/D_N$. 
The $\mu_S$ and $k_{FS}$ act here as convenient arbitrary 
normalizations. 
These peaks are superimposed on the straight lines 
that would represent the DOS in the $D_N \rightarrow \infty$ limit.
If one normalizes, as we do,
the energies in terms of $\mu_S$, and the DOS as explained
above, then the slope of these straight lines would be $\pm 1$. 
Our results are shown in Fig.~\ref{f30}.
The top panels ((a) and (c)) show
the  DOS (calculated from and
normalized as in Eq.~(\ref{dos3})) over a broad energy range. 
Panel (a) is for undoped graphene ($\mun=0$)
and panel (c) corresponds to a relative doping of $\mun = 0.2$.
For these two cases,
there are four curves shown that correspond
to four different graphene widths (see caption).
The Thouless peaks are at their predicted
positions. Their magnitude tends to decrease as the width increases, 
with the results for largest width, $D_N = 600$, approaching 
the signature linear dispersion for bulk graphene.
With the introduction of doping (panel (c))
the results are shifted, in normalized energy units, by $-\mun$ 
from the Dirac
point, resulting in the shifting of the gap away from zero energy 
for all widths shown.
To more clearly discern the Thouless peaks, the bottom panels, 
which correspond to the same parameter values
as the top ones,  illustrate 
the DOS over a smaller energy range.
In Figs.~\ref{f30} (b) and (d), the Thouless peaks in the DOS are clearly seen to
occur at energies that coincide
with the expression discussed below 
Eq.~(\ref{thou}): for $D_N = 300$, the first peaks arise at 
energies corresponding to $|E_c/\mu_S| = \pi/300 \approx  0.01$,
while for $D_N = 600$, we have, $|E_c/\mu_S| =\pi/600 \approx  0.005$. 
In Fig.~\ref{f30}(b), the curves representing the smaller
widths, and correspondingly larger $E_T$,
are absent since (as panel (a) shows)
they emerge beyond the given energy window.
These results illustrate also the precision and reliablity of
our methods. 

We now return to the \sns trilayer
and investigate the roles that both the Thouless and
superconducting energy scales play by considering
the local DOS
of a \sns nanostructure in both the S and N regions.
After inserting the quasiparticle expansions found in Eq.~(\ref{expand}),
the general expression for the DOS in 
Eq.~(\ref{dos}) can be
rewritten as,
\begin{align}
\label{dos2}
\dfrac{N(x,\epsilon)}{N_0(\mu_s)} 
&=
\frac{\mu_S}{T k_{FS}d }\int_0^{k_c} d\widetilde{k}_y   
\Bigl[\Bigl(\Bigl|\sum_{n,q}  s_{n,q}\sin(k_q x)\Bigr|^2 \nonumber \\
&+ 
\Bigl|\sum_{n,q}  t_{n,q}\sin(k_q x)\Bigr|^2 \Bigr) \cosh^{-2} 
\Bigl(\dfrac{\epsilon-\epsilon_n}{2T}\Bigr) \nonumber \\
&+\Bigl(\Bigl|\sum_{n,q}  w_{n,q}\sin(k_q x)\Bigr|^2 
+ \Bigl|\sum_{n,q}  z_{n,q}\sin(k_q x)\Bigr|^2 \Bigr) \nonumber \\
&\times 
\cosh^{-2}\Bigl(\dfrac{\epsilon+\epsilon_n}{2T}\Bigr)
\Bigr],
\end{align}
In calculating the DOS for the \sns cases, 
we take the  eigenvectors and eigenenergies, 
self-consistently calculated as explained above,
and insert them into Eq.~(\ref{dos2}). 
When $\widetilde{\omega}_c <1$, the case considered here,
the relationship between the bulk transition temperature $T_0$
and $\Delta_0$ is found using Eq.~(\ref{lam}),
together with,\cite{bogobook}
\begin{align}
\lambda^{-1} = \int_0^{\omega_c} \dfrac{d\xi}{\xi} \tanh[\xi/(2 T_0)],
\end{align}
which results in the weak coupling
limit in the BCS 
relation\cite{cite:sonin1} $\Delta_0=(\pi/\gamma_E)T_0$,
with $\gamma_E$ being the Euler constant.
We take the  low temperature limit 
$T/T_0 \approx 0.016$.  This is the same temperature 
as in the  previous plots. 
The energy-resolved DOS is then determined at two locations:
one at the middle of one of the superconducting regions, 
and the other at the 
center of the sample (normal region). In these plots, we will  
normalize the energy (measured as usual from the chemical
potential) by $\Delta_0$, and the plotted DOS (as before) by $N_0(\mu_S)$. 
Thus, if our 
plots were performed for an infinite normal sample, they would
of course still be straight lines but the slope would 
now be  $\pm \Delta_0/\mu_S$, which is the same
as $1/\Xi_0$ via the relations mentioned in Sec.~II. 

There are now two energy scales to consider. One is, as 
ordinarily in all
superconductors, the  bulk gap $\Delta_0$. In addition, because our
system is two dimensional, has a linear, massless 
dispersion relation, and is
finite in the $x$
direction, the DOS is also, as we have seen, drastically affected 
by the Thouless energy.\cite{saito} 
The quasiparticles in the previously discussed
graphene nanostrip were confined solely by the two outer boundaries.
Now due to the intrinsically superconducting regions,
there is also possible partial confinement 
by the self consistent pair potential $\Delta (x)$,
which due to Andreev scattering events (normal or specular), 
leads to a modification
of the relevant energy scales associated with the peaks 
in the quasiparticle spectra. Thus, the two scales interact.
For large N graphene 
widths ($d_N \gg \xi_0$),
the normalized energy spectrum has the gap set primarily by 
the Thouless characteristic  energy, 
and a peak structure\cite{cite:been2} at multiples of $E_T$. 
In the \sns geometry the relevant scale is now\cite{xtra2} $E_c= (\pi/2) E_T$.
It follows 
then that 
$E_c/\Delta_0=(\pi \Xi_0)/(2 D_N)$. 
The interplay  between this 
ratio and 
that of $\Delta_0$ to $\mu_S$, 
can for a given energy range, result in many  
additional resonance
peaks.
In the non self-consistent treatment, these peaks occur exactly at integer multiples of $E_T$,\cite{cite:been2} 
but as will be seen below, self consistently accounting for
proximity effects in our finite sized system can modify this picture. 
Hence, as a consequence of the existence of the Thouless scale,
the behavior of the local DOS in a \sns heterostructure is 
strongly dependent on the size of each region.
If however the normal graphene channel is much narrower than $\xi_0$, 
the lowest energy scale is $\Delta_0$.

To illustrate these issues, results for  $D_S=150= 1.5 \Xi_0$ and 
various values of $D_N$  are exhibited in Fig.~\ref{f3}.
These include cases where $d_N<\xi_0$ and cases where
it is larger, thus  demonstrating the relative
relevance of both the $\Delta_0$ and $E_T$ energy scales in
different situations. 
Thus, consider first the bottom panels (c) and (d), in this 
Fig.~\ref{f3}, where results for the N region are plotted. In panel (d)
we have $d_N=3 \xi_0$ ((red) solid curve)
and $d_N=6 \xi_0$ ((green) dashed curve)
so that the influence of the S portions of the
sample is, while as we have
seen not negligible, weak. One clearly sees
that the results can still 
be described by a straight line of slope $1/\Xi_0 =0.01$
on which there are superimposed peaks and oscillations related
to both the $\Delta_0$ and the $E_T$ scales. For the parameter values in this
panel we have (see discussion above) 
that when $\epsilon/E_c=1$ then 
the normalized energy $\epsilon/\Delta_0$ is about 1/2 for $d_N=3\xi_0$ 
and half that for the other case shown. One sees indeed this behavior
in this panel (d). On the other hand, in panel (c), where results for
smaller values of $D_N/\Xi_0$ are shown, the influence
of the Thouless scale is very weak for $d_N=\xi_0$ ((blue) dashed curve) 
when the two energy scales roughly coincide)
and nearly nonexistent when $d_N < \xi_0$ ((black) solid curve).

The results in Fig.~\ref{f3} for the local DOS in the S region are 
very different. These are shown in panels (a) and (b) for the same
values of $D_N$ and the same plotting conventions as for
panels (c) and (d) respectively. We now see a very clear
energy gap close to the bulk value. The influence of the
Thouless energy is reduced now to some weak additional peaks
at higher energy. We can see that at the energy scales shown the
effect of the bulk normal state linear DOS is not visible, although
of course this is an artifact arising from the energy range
plotted and the increasing behavior reappears eventually at larger values of 
$\epsilon \gg \Delta_0$. Even though $\mun = 0$, 
particle-hole symmetry breaks down in the N
region. 



The local DOS is 
very dependent on the doping level. To show 
this we
display in Fig.~\ref{f4} results for the DOS at $\mun =0.2$.
In this case, the DOS for bulk ($D_N \rightarrow \infty$)
normal graphene (zero Thouless energy and zero pair potential)
is still (see Fig.~\ref{f30}) 
a straight line  but with the origin shifted.
For our parameter values, and indeed for any reasonable
parameter values in our context, this origin is shifted out  of the horizontal
scale 
in the energy ranges of order $\Delta_0$ shown in this figure.
The four panels in the figure are arranged exactly as those in Fig.~\ref{f3}
and correspond (both the panel arrangement and the (color or)
structure of the curves) to exactly the same cases. One can
see that when the doping amount changes,
this DOS becomes
more complicated because of changes
in the quasiparticle bound states  
shifting with the
Fermi level. Thus, the results for large $D_N$ in the N region
(panel (d)) show now only a faint trace of any gap, either superconducting 
or Thouless: the DOS is nearly linear at small energies, 
possessing a V shape at the Dirac point. This subgap structure
is a modification to the traditional Andreev bound states\cite{cite:james} that arise 
in the spectrum of conventional superconductor-normal metal systems.
In panel (c), when the 
$d_N$ is comparable  to the correlation length, the ``V" behavior
still persists (dashed (blue) curve) but it is completely gone, and
replaced by a gap, when the thickness is below $\xi_0$ (solid (black)
curve). The panels (a) and (b), corresponding to the S region,
are less strikingly different from the corresponding ones in Fig.~\ref{f3}
but they do show an intriguing additional structure in the gap region.

\begin{figure}
\centering
\includegraphics[width=.4\textwidth]{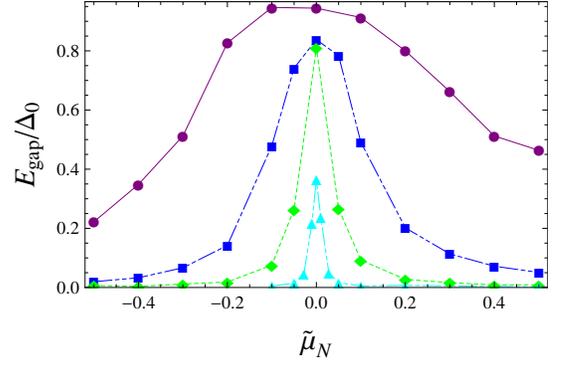}
\caption{(Color online). The excitation gap, $E_{\rm gap}$ as
defined in the text, 
as a function of the relative doping parameter, $\mun$. Four different 
normal graphene widths are considered:
$D_N = 20, 50, 100$, and $300$ ((red) circles, (blue) squares,
(green) diamonds, and (cyan) triangles respectively. 
Lines are straight segments joining points.
}
\label{f5} 
\end{figure}

In view of the strong effect, as evidenced in the 
comparison of Figs.~\ref{f3} and \ref{f4}, of $\mun$ on the
gap structure in the DOS
it is interesting to further examine in a more direct
way 
the induced gap in the quasiparticle spectrum.
This we do by extracting from our numerical results
the  eigenvalue from the self consistent spectra
obtained from Eq.~(\ref{eq:mat}) for which $\epsilon_n$ 
as measured from the chemical potential is minimum.
We call this quantity the excitation gap, and
denote it by $E_{\rm gap}$,
which is generally determined by  
longitudinally directed (along $x$) 
trajectories corresponding
to small $k_y$. 
The results are shown 
in Fig.~\ref{f5} where we show the evolution of $E_{\rm gap}$
normalized by $\Delta_0$ (so that the quantity plotted is
non-negative  and less than unity)
as a function of $\mun$. Results for four different
values of $D_N$ are shown, encompassing values both above
and below $\Xi_0$: $D_N=20=0.2 \Xi_0$ (circles), 
$D_N=50$, (squares), $D_N=100$ (diamonds),
and $D_N=300$ (triangles). In all cases $D_S=150$. 
The range of $\mun$ and values of $D_N$
that result in a gap are of course consistent with the DOS results above.
The results shown illustrate that structures including 
narrower normal graphene layers possess energy gaps that are 
much more robust to
changes in N layer doping. 
The contraction of the gap
with increasing $D_N$
is qualitatively similar to what is observed
in conventional three dimensional systems,\cite{gap}
but, as mentioned above, the structure of the gap
amplitude and of the DOS is very different.

\begin{figure}
\centering
\includegraphics[width=.48\textwidth]{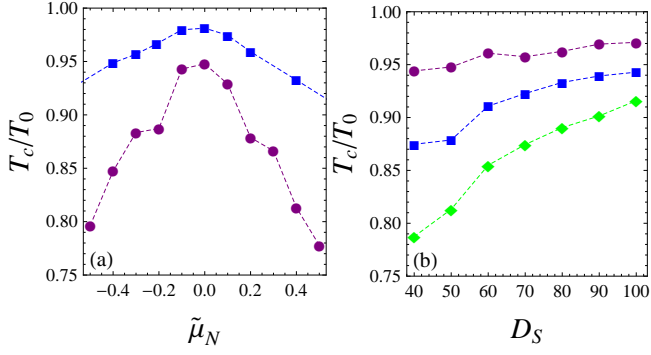}
\caption{(Color online). The critical temperature $T_c$
(normalized by $T_0$)
for
a \sns system as a function of (a) $\mun$,
and (b) the S width, $D_S$. 
In (a) the doping dependence is shown for two values of
$D_S = 50$ ((red) circles) and $100$ ((blue) squares).
In (b), results are shown as 
a function of $D_S$ for three doping levels:
$\mun = 0$ ((red) circles), 
$0.2$ ((blue) squares) and $0.4$ ((green) diamonds). The  
doping level in these cases
has a dramatic effect on $T_c$ for small $D_S$,
but
is less detrimental for
larger $D_S$, where the curves tend to coalesce 
in the limit of bulk S widths. 
In both (a) and (b),
the normal graphene layer has $D_N = 100$. 
}
\label{f6} 
\end{figure}

Up to this point, we have considered the  
low temperature limit. It is of interest both 
experimentally and theoretically to now
turn our attention to the calculation of the
critical temperature $T_c$ of the \sns structures and its dependence
on doping levels and geometrical parameters.
This quantity is calculated using the efficient eigenvalue method  described
by Eq.~(\ref{tc}) and the discussion below it.
Results are presented in terms of the ratio $T_c/T_0$ and displayed
in Fig.~\ref{f6}. In the left panel, results are given
as a function of relative doping level $\mun$ for two values of $D_S$: 
$D_S=\Xi_0=100$ ((blue) squares) and $D_S=\Xi_0/2$ ((red) circles). We keep 
$D_N=\Xi_0$ constant in this figure. We see than increasing $|\mun|$ 
to moderate values, that
is,  
decreasing the Fermi level mismatch, leads to (Fig.~\ref{f6}(a))
a reduction in $T_c$ via the corresponding
increase in the interlayer coupling.
This effect is more pronounced
for thin S layers, where $T_c$ can vary with $\mun$ in a nontrivial fashion.
It is remarkable, however, that $T_c$ remains rather high even when
$D_S$ is smaller than the correlation length. This is is stark contrast to 
ordinary three dimensional \sns systems, where $T_c$ drops much more rapidly for small
thicknesses satisfying 
$d_S \lesssim \xi_0$. There is also a clear asymmetry in the critical temperature as a function
of doping, where for a given magnitude $|\mun|$,
electron doping more strongly reduces $T_c$.
In the right panel, 
(Fig.~\ref{f6}(b)) we illustrate that varying the width of the superconductors
has a considerably greater impact on $T_c$ for moderate values of $\mun$
than when the mismatch is large. In the latter case the $D_S$ dependence remains
weak as long as $D_S$ is still comparable to $\Xi_0$, however (and consistent with panel (a)) 
the superconducting regions that have widths a fraction of 
the coherence length reveal the richest behavior. 
Continuing to reduce $D_S$ beyond some critical value, of course
results in the graphene system eventually becoming nonsuperconducting,
as Cooper pair formation is inhibited. 

\begin{figure} 
\centering
\includegraphics[width=.48\textwidth]{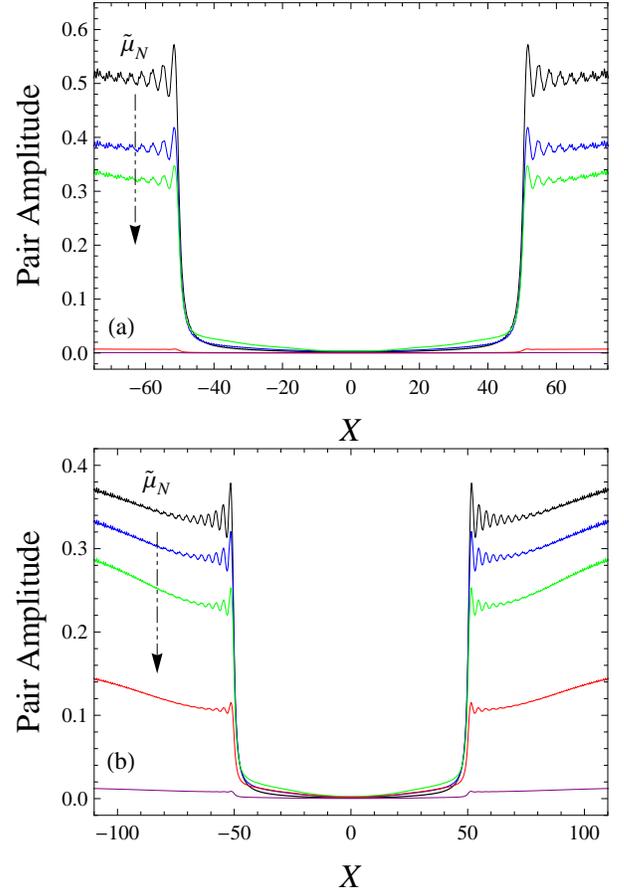}
\caption{(Color online). The normalized pair amplitude as
a function of position. 
In (a), the temperature is
set at $T = 0.87 T_0$, 
and the normal and superconductor  widths are
$D_N = 100$ and $D_S = 50$ respectively.
In (b), we have $T = 0.92 T_0$, while $D_N =100$ and $D_S = 100$.
In both cases,
the  doping parameter, $\mun$,
is varied from $0$ to $0.4$ in increments of $0.1$. The arrow depicts the
direction of increasing $\mun$ which eventually leads to
the vanishing of the pairing correlations.
}
\label{f7} 
\end{figure}
Results such as those shown
in  Fig.~\ref{f6}(a) imply that at a fixed temperature,
variations in the doping parameter, $\mun$,
can lead to a \sns system transitioning from
a superconducting state to normal one and vice versa.
Since graphene doping can be effecttively 
tuned via application 
of an external electric field,\cite{cite:novoselov1,ahn} 
this may offer possibilities as a carbon-based \sns switch 
for supercurrent flow. This question is, therefore, worthy
of further discussion. 
We thus expand on this point by
showing in Fig.~\ref{f7} the normalized pair amplitude  $F(X)$
plotted as a function of $X$ 
for several positive values of $\mun$. 
Each panel is at a different fixed temperature and
the geometrical parameters,  $D_S = \Xi_0/2$ and $D_S =\Xi_0$
are chosen to correlate respectively
with the (red) circled and (blue) squared data of Fig.~\ref{f6}(a).
The two representative temperatures that we investigate
are $T=0.87 T_0$ (panel (a)), and $T=0.92 T_0$ (panel (b)).
The graphene region that is
intrinsically nonsuperconducting
has a width in both cases corresponding to $D_N = \Xi_0$.
One can see in Fig.~\ref{f6}(a) that for 
the smaller $D_S = \Xi_0/2$, the temperature $T=0.87 T_0$ 
corresponds to $T_c$ 
near $\mun \approx 0.35$, and  
for $D_S = \Xi_0$, the temperature $T=0.92 T_0$
results in $T_c$ near $\mun \approx 0.4$.
The
regions for which positive $\mun$ is smaller being 
superconducting (the corresponding negative $\mun$ differ slightly due to the
electron-hole doping asymmetry). This is more clearly seen in Fig.~\ref{f7},
where as $\mun$ is increased, the pair amplitude is seen
to decrease before plummeting abruptly 
to zero as $\mun$ reaches its critical value: near
$\mun \rightarrow 0.35$ in panel (a) or near $\mun \rightarrow 0.4$ for panel (b). 
Thus, if this transition can be
manipulated via electric fields, abrupt switching will result.

\section{Conclusions}

We have studied in this paper the proximity effects that
occur in clean, doped and undoped, graphene-based \sns trilayers. We
have created and implemented a fully self-consistent procedure to calculate
the electron and hole wavefunctions and energy spectrum of the system, from which we have extracted
the pair amplitude and the local DOS. 
We also developed a semianalytical
and computationally efficient 
linearized method
that can calculate
the transition temperature, $T_c$, of the system.

We have found that the behavior of the pair amplitude
near the interfaces (the directly
observed proximity effect) depends
strongly on the relative doping levels of the S and the N 
portions, and that 
the pair amplitude is described by two different
length scales. 
One length scale is related to penetration of the superconducting correlations,
and is long ranged (relative to $\xi_0$) and the other scale is short ranged,
and correlates to Cooper pair leakage from the S regions near the interfaces.
We illustrated that if the normal 
graphene layer is weakly doped,  
specular Andreev reflection can lead to
superconducting correlations  penetrating into the normal graphene region.
The local DOS exhibits
a number of striking features, arising from the interplay between
the superconducting and the Thouless energy scales. This interplay
depends of course on geometry, where
the two energy scales overlap when the
graphene layer widths are on the same order as $\xi_0$.
For our larger structures 
(with widths exceeding $\xi_0$), undoped normal regions
revealed resonant peaks and energy gaps at characteristic energies proportional to
the Thouless energy scale $E_T$. By moderately doping the N region,
there was an emergence of Andreev bound states in the S regions and
a destruction of the energy gap.
The smaller \sns structures (with widths smaller than $\xi_0$)
revealed energy gaps that are linked mainly to the $\Delta_0$ scale,
and are more robust to doping.

We also developed a general microscopic
method for calculating $T_c$ for  \sns nanostructures,
by linearizing the DBdG equations and the self consistency condition.
We found that for small S layer widths,
decreasing the Fermi level mismatch leads to 
a nontrivial reduction in $T_c$.
The critical temperature also exhibited
a clear asymmetry as a function
of doping, and typically electron doping had a greater impact on reducing $T_c$. 
Thus if doping is to be modified by an electric field,\cite{cite:novoselov1,ahn}
the polarity\cite{shin} of the field can have an important effect on the critical temperature.
The study of $T_c$ revealed reentrant behavior
as a function of doping. These behaviors may lead
to switching phenomena as a function of applied electric
field, and thus depending on the bias, superconductivity can be turned on or off.
The effectiveness of graphene 
as a low temperature field effect device therefore
depends in large part by
the proximity effects,
which can only be accounted for within a  self-consistent framework.
This work represents the first step, a proof of principle, as to
the use of our self consistent methods in graphene. Other
issues, such as those
related to ferromagnetically doped graphene in contact
with a superconductor region, can also be examined using the
same techniques. We expect that many aspects of the ever
intriguing behavior of graphene-based heterostructures will
be illuminated via application of these methods.

\acknowledgments
This work is supported in part by ONR and by grants
of HPC resources from 
DOD (HPCMP) and from the Minnesota Supercomputer
Institute. M.A. wishes to thank J. Linder for conversations. 


\end{document}